# Proposed STAR Time of Flight Readout Electronics and DAQ

J. Schambach for the STAR Collaboration TOF Group
*The University of Texas at Austin, Austin, TX 78712, USA*

The Time-of-Flight system is a proposed addition to the STAR detector currently installed at RHIC. The proposal is based on a new technology called the Multi-gap Resistive Plate Chamber (MRPC), and consists of approximately 23,000 channels of detectors distributed over 120 trays. Each TOF tray consists of 192 detector channels and three different types of electronic circuit cards, called "TFEE", "TDIG", and "TCPU", listed in order of the data flow. Every fourth tray also contains a "TMIT" card that transmits the data over a fiber to the TDRC which is part of STAR DAQ. The TFEE contains the analog front end electronics. The output of TFEE is passed to the TDIG, where the data are digitized (using the CERN HPTDC ASIC). The TCPU formats and buffers the digital detector information. This formatted data is passed to TMIT, which transmits it over an optical fiber to the data receiver TDRC in the STAR DAQ room. The architecture of this readout chain and DAQ will be described, and first results from prototypes of the component boards will be discussed.

## 1. THE TOF PROPOSAL

The main goal of the relativistic heavy ion program at RHIC is to produce a new form of matter, the Quark-Gluon Plasma (QGP), and to study Quantum Chromo Dynamics (QCD) in matter at high temperature. Initial experimental measurements of particle multiplicity and transverse momentum (energy) distributions indicate that the high energy density being reached in nucleus-nucleus collisions at RHIC is unprecedented. Recent RHIC results have offered a new view into nucleus-nucleus collisions with respect to studying the gluon dominated initial phase, partonic and hadronic evolution dynamics, and hadronic freeze-out scenarios.

A unique strength of the STAR detector at RHIC is its large, uniform acceptance capable of measuring and identifying a substantial fraction of the particles produced in heavy ion collisions. This large acceptance is central to STAR's scientific capability and has already resulted in new and intriguing physics results like the recent measurement of the suppression of back-to-back jets in central Au+Au collisions. Large acceptance detectors central to the STAR heavy ion physics program are the Silicon Vertex Tracker (SVT), the Time-Projection Chamber (TPC), and the Barrel Electromagnetic Calorimeter (BEMC), all having an acceptance covering $2\pi$ in azimuthal angle and $|\eta|\leq\sim1.5$ in pseudo-rapidity. STAR proposes a barrel Time-of-Flight (TOF) detector based on "Multigap Resistive Plate Chamber" (MRPC) technology, matching the acceptance of these detectors [1]. This upgrade will provide essential particle identification capability. Specifically, the percentage of identified kaons and protons will double to more that 95% of those produced within the acceptance of the TOF barrel ($|\eta|\leq\sim1$), greatly enhancing the discovery potential of STAR. This increase in particle identification efficiency over a large solid angle is especially important for measurements of multi-particle correlations since the feasibility of such measurements depends (on average) on the single particle efficiency raised to the power of the number of particles used in the correlation. The extended momentum range for particle identified spectra provided by the MRPC barrel TOF detector is crucial to understand the information contained in the large scale correlations and fluctuations being observed in Au-Au collisions at RHIC.

## 2. THE DETECTOR

Large acceptance TOF coverage has been an integral part of the proposed STAR detector since its inception. At that time, the TOF detector technology proposed was based on mesh-dynode phototubes and was deemed to be too expensive. The development of new, low-cost MRPC technology for the ALICE experiment combined with the parallel development of new electronic chips at CERN provides a cost effective means to fulfill the physics-driven requirements of the original STAR design. After extensive testing of MRPC technology at CERN and BNL in addition to its successful implementation in the HARP experiment, this technology is mature; a full-scale 168 channel MRPC prototype tray has been installed in STAR for in-situ testing during the 2003 RHIC run.

Figure 1 shows two side views of an MRPC module appropriate for STAR. The upper (lower) view in this figure shows the long (short) edge of a module. The two views are not shown at the same scale. An MRPC is a stack of resistive glass plates with a series of uniform gas gaps. Electrodes are applied to the outer surface of the outer plates. A strong electric field is generated in each sub-gap by applying a high voltage across these external electrodes. All the internal plates are electrically floating. A charged particle going through the chamber generates avalanches in the gas gaps. Because the glass plates are resistive, they are transparent to charge induction from avalanches in the gaps. Typical resistivity for the glass plates is on the order of $10^{13}$ $\Omega$/cm. Thus the induced signal on the pads is the sum of possible avalanches from all gas gaps. The electrodes are made of resistive graphite tape and are also transparent to charge. Copper pickup pads are used to read out the signals. The graphite electrodes have a surface resistivity of $10^{5}$ $\Omega$ and cover the entire active area. The outer and inner glass plates are 1.1 mm and 0.54 mm thick, respectively. They are kept parallel by using 0.22 mm diameter nylon fishing line as a spacer. The signal is read out by an array of copper pickup pads. The pickup pad layers are separated from the outer electrodes by 0.35 mm of Mylar.

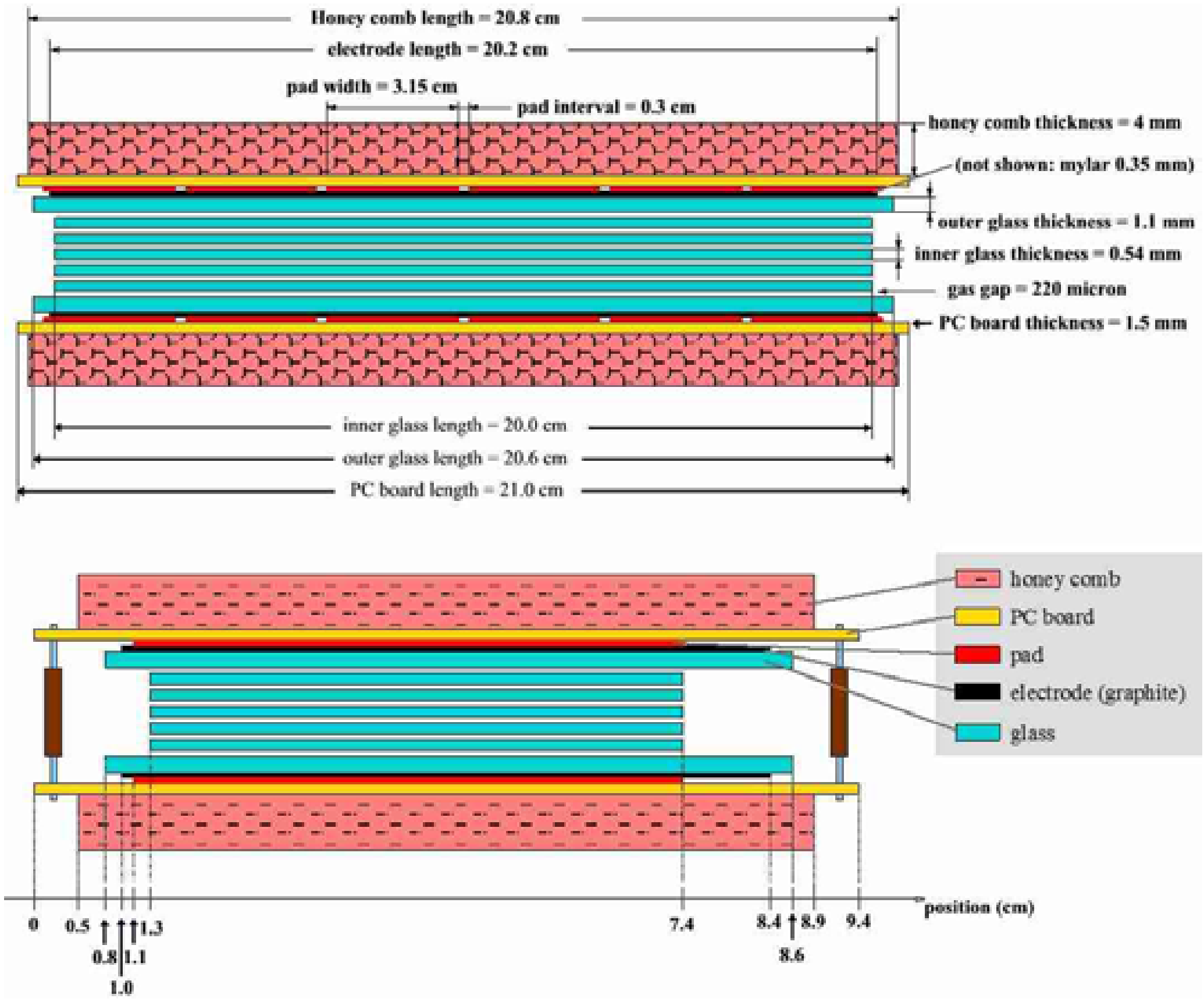

Figure 1: Two side views of the structure of an MRPC module. The upper (lower) view shows the long (short) edge. The two views are not shown at the same scale.

The primary mechanical structure of the system is a “tray.” These hold the detectors in three dimensions at specific positions and angles, support the on-detector electronics, and form all but one wall of the gas volume. A tray is an aluminum box with 50 mil (0.13 cm) wall thickness and welded corners. It takes 120 trays to cover the cylindrical outer radius of the TPC. The trays are arranged as two adjoining cylindrical shells of 60 trays each. Each cylindrical shell subtends approximately 1 unit of pseudo-rapidity. The detectors inside the gas volume are standard MRPC “modules,” 6 channels each, with detector channel dimensions of 3.3cm×6.1cm. It takes 32 modules to fill a tray. The “top” of the tray is a custom 1/8” thick electronics board called the TOF Front End Electronics (TFEE) plate. These TFEE plates are simultaneously the “top” of the gas box, the signal feed-through mechanism, the actual FEE boards, and the mounting location of the TOF Digitization (TDIG) boards just above the TFEE boards. Thus, the electronics exist as two layers, the TFEE boards themselves are both the first layer of electronics and the tray top. Mounted above these is another layer of boards, TDIG, that perform the digitization. At the high-z end of the tray are two more electronics boards, “TCPU” and “TMIT” (described later), which perform the communication of the digital data. A gas system provides a mixture of 90% Freon (R134a), 5% Isobutane, and 5% SF6 to the TOF chambers at the correct pressure.

## 3. ELECTRONICS DESIGN

A TOF system measures time intervals, which are defined by independent electronic measurement of one “start time” and some number of “stop times” in each experimental event. In the proposed system, we perform the timing digitization on-detector for about 23000 channels, with timing relative to a common clock to determine the stop time for a particle's flight. The start time, or time of the collision, is determined by the vertex position detector (VPD), scintillator-PMT detectors close to the beam pipe, which is digitized in the same way as the stop times. This information (and pulse width data for time-walk correction) is transmitted digitally to the STAR DAQ for timing analysis:

$$TOF = t_{stop} - t_{start}$$

The total resolution after all corrections must be ≤100ps to achieve STAR's physics goals. For a start resolution of «50ps attained from the VPD, and a 30ps contribution from the timing corrections, a pure stop resolution of less than 80ps is dictated. The MRPC detectors developed for this project have demonstrated pure stop resolutions in the range from 50-70 ps using CAMAC TCD's with 50ps time bins and prototype front-end electronics very similar to that described later in this section.

The other basic performance requirement of the TOF electronics is to provide sufficient data buffering and transport capacity to deliver the TOF data to the STAR DAQ, within the constraints of hit rates, trigger (event) rates, and trigger latency.

Shown in Figure 2 is the top-level diagram of the electronics for the start and stop sides of the present system. The electronics chain on the start side is very similar to that on the stop side by design. The individual electronics boards seen in this figure are discussed in detail below.

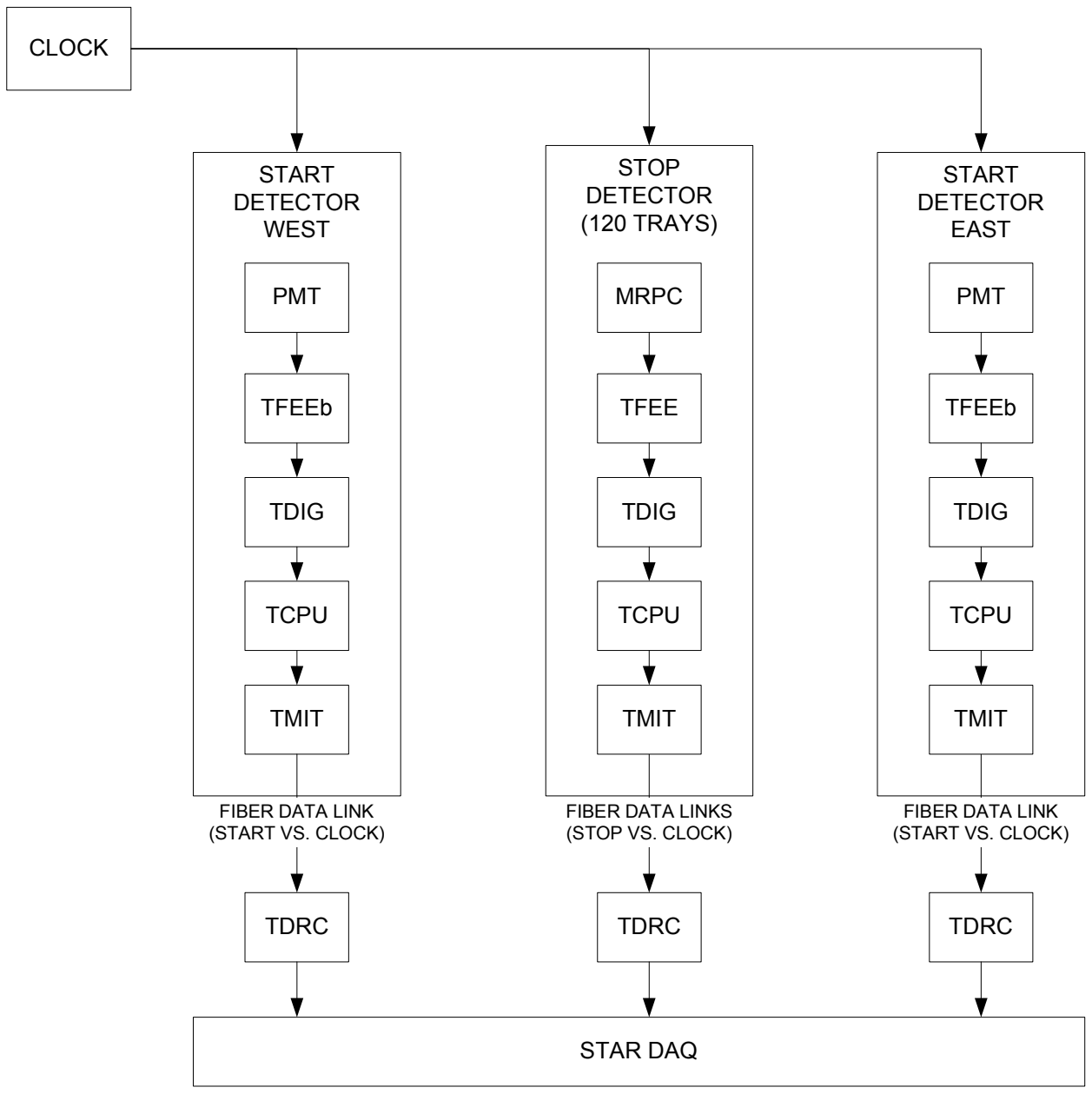

Figure 2: Top level electronics design

The approach for the TOF readout electronics design uses on-detector electronics on the MRPC detector trays, with tray-level data buffering, and fiber-optics data communication with STAR DAQ. The complete STAR TOF system will consist of 120 trays, along with their support hardware (power supplies, cooling, etc.) Each tray includes: 192 MRPC detector channels, 8 amplifier / discriminator circuit cards ("TFEE"), with 24 channels each, 8 time-to-digital converter cards ("TDIG"), with 24 channels each, and 1 CPU card ("TCPU") to store and forward data for all 192 data channels. Every fourth tray also contains a data communications card called TMIT. The TMIT card transfers the data to the DAQ receiver card ("TDRC").

All the detector electronics functions are partitioned between 4 different circuit cards that are located on the detector trays and one card in the STAR DAQ system. Each MRPC module contains 6 detector channels. Four modules feed into one TFEE. Each TFEE feeds one TDIG. The 8 TDIG cards in a tray feed into a single TCPU. Data from 4 TCPU's will feed into a single TMIT located as a daughterboard on one of the TCPU's.

The TMIT card sends its data to a TDRC located in the DAQ room via optical fiber. The TDRC serves as the receiving end of the fiber and the interface to the STAR DAQ system. In the current design the TDRC will be implemented as a VME card, since most of the current STAR DAQ system is based on VME.

The partitioning of the electronics on a tray into these circuit cards and the communication between these cards and interfaces to other STAR systems is shown schematically in Figure 3.

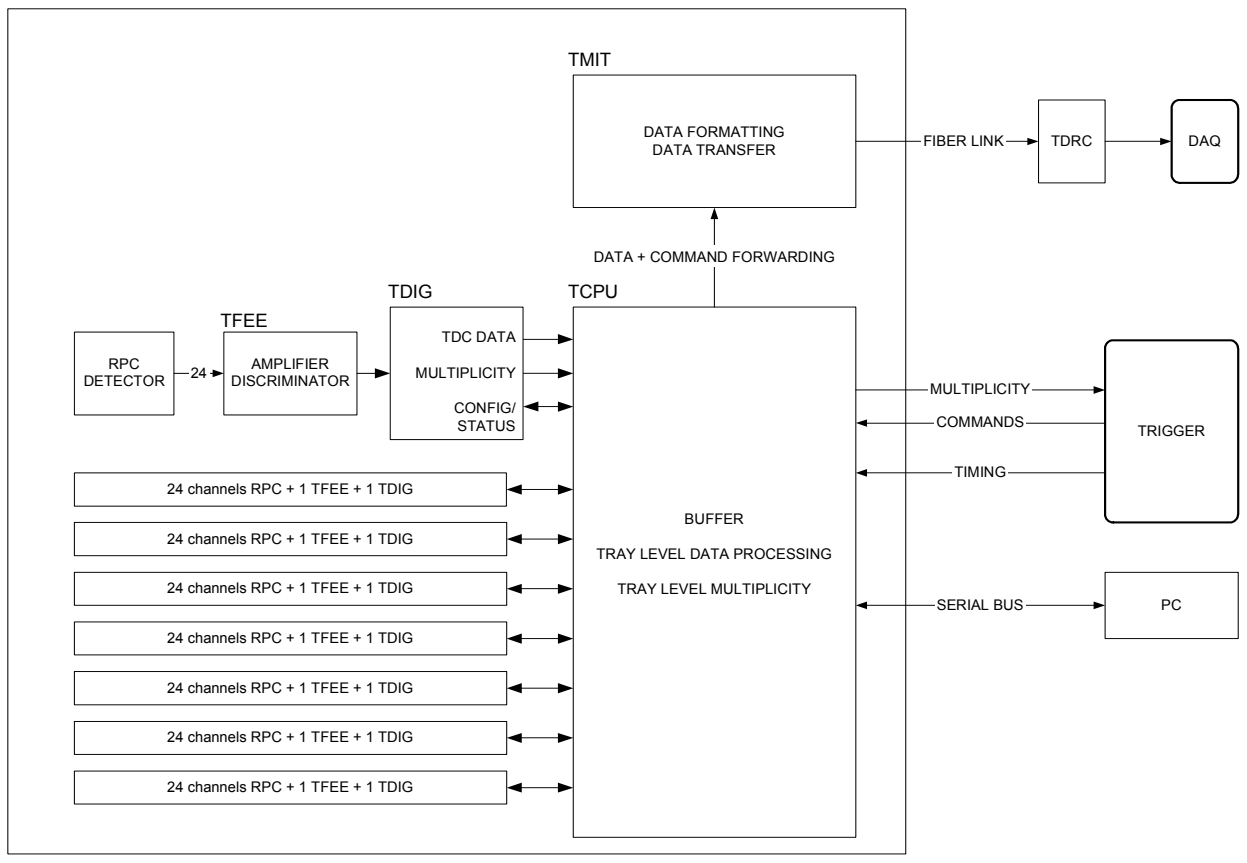

Figure 3: Electronics and communication at the tray level

The following subsections describe the various circuit cards.

## 3.1. TFEE

This card contains pre-amplifier and discriminator circuits to accommodate 24 pad signals from 4 MRPC modules. A TOF detector tray will have 8 TFEE cards. The principal functions of the TFEE and TDIG are shown in Figure 4.

The preamp device, a Maxim 3760, is a low noise input trans-impedance integrated circuit whose gain and rise time characteristics are well-defined by internal feedback. This component is commercially available for use as a photodiode receiver preamp in data communication applications. Several designs employing this chip have shown excellent timing performance when connected to actual MRPC pads - the Maxim 3760 has been used extensively for the past two years by both the STAR and ALICE TOF groups.

An ultra-high speed integrated circuit comparator, the AD96685, serves as a simple leading-edge discriminator with externally controlled threshold. This circuit has also been used successfully in prototype TOF systems at STAR. The output pulse width, i.e. the "Time Over Threshold," will be used to infer the input signal rise time for the slewing correction.

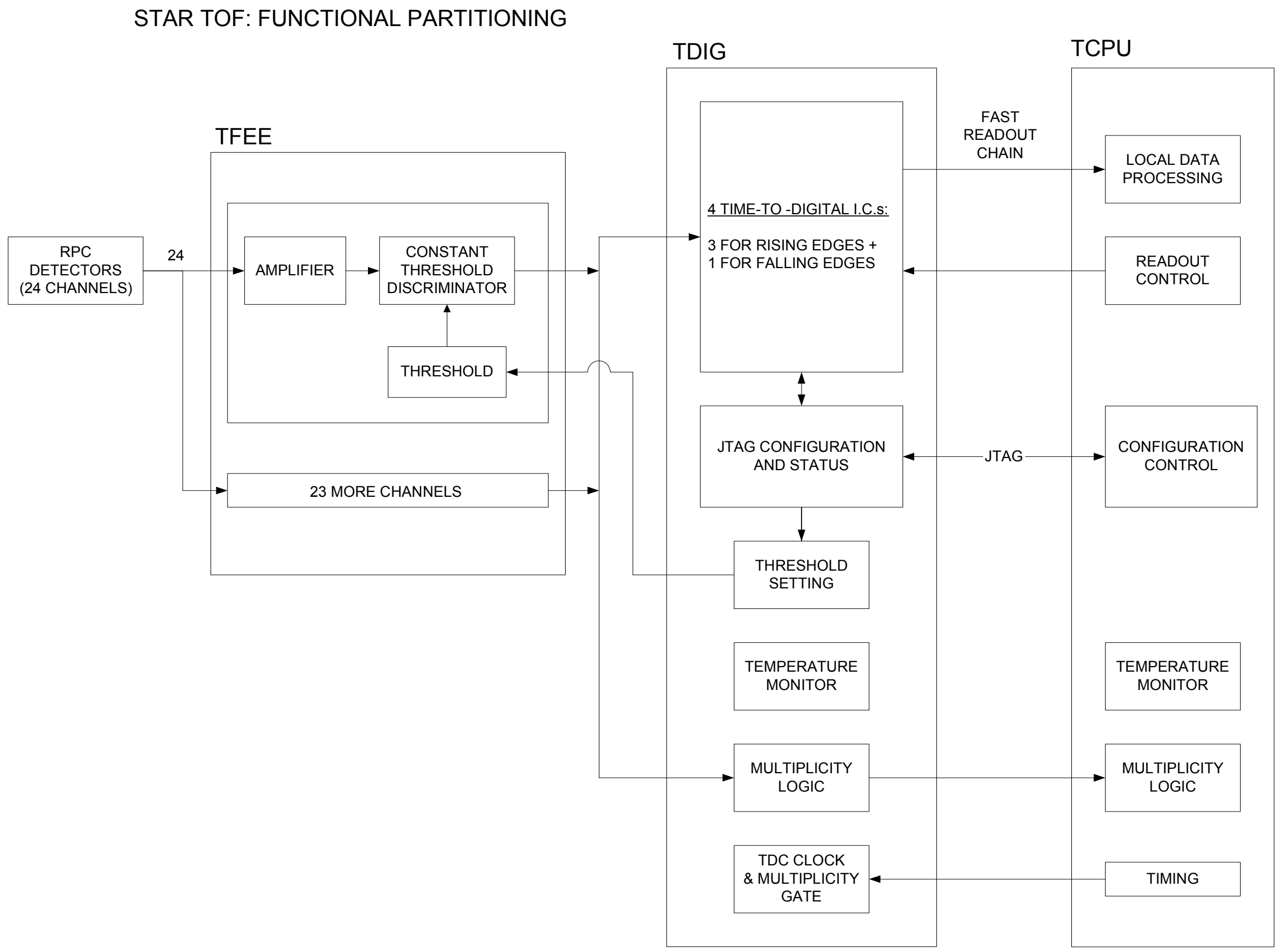

Figure 4: The front end electronics (TFEE) and digital sampling (TDIG) circuit cards.

The TFEE is an integral part of the tray mechanical assembly. This results in a short signal path and maximum shielding. The TDIG cards are stacked above their respective TFEE card to reduce RF cabling. Multiplicity sums have dedicated parallel data paths from each TDIG card to the TCPU. Provisions will be made during the design of the TFEE cards to optionally bypass the preamp section. This will allow these boards to operate when the inputs are photomultiplier signals, not MRPC signals. This will allow the identical signal processing chain to be used on the start side with no additional board designs.

## 3.2. TDIG

The TDIG card measures leading and trailing edge timing for 4 MRPC modules (24 detector channels). These cards are mounted directly on the TFEE cards – one TDIG card per TFEE card. The discriminator signals, clock, multiplicity gate, and L0 trigger readout commands are primary inputs while the hit edge timing data and 5-bit partial multiplicity sums are outputs. The calculation of the pulse width for the time-over-threshold based slewing correction, and any resultant data formatting, will occur downstream of this card.

The multi-channel “High-Performance TDC” (HPTDC) ASIC developed at CERN for the ALICE and CMS experiments [2, 3, 4] is our first choice for the TDC measurement. In addition to meeting our time resolution requirements, it has efficient and flexible triggering and readout features. The trigger matching function allows acquired data to be read out from the built-in buffer in an order that accounts for trigger latency. As a backup, we will evaluate a device under development at the University of Oulu in Finland. The Oulu device as specified has adequate timing resolution but does not include readout features of the CERN HPTDC. However, the Oulu TDC is a full custom CMOS ASIC achieving 20 ps timing precision with very low power consumption of < 5 mW/channel. NASA is currently evaluating chips they have received from Oulu.

Leading edge timing for 24 channels with 25 ps binning will be provided by 3 HPTDC devices operating in 8 channel, “Very High Resolution” mode. The trailing edge times of all 24 channels will be determined by 1 device operating in 32 channel “High Resolution” mode with 100 ps time bins. A built in hardware handshake protocol allows 16 HPTDC devices to share a 80Mbit/sec serial output port so that only 2 data cables are required per TOF tray (4 TDIG boards, 96 MRPC pads each).

Additionally, this card will have slow serial interfaces for TDC and logic configuration, control of discriminator threshold and temperature monitoring.

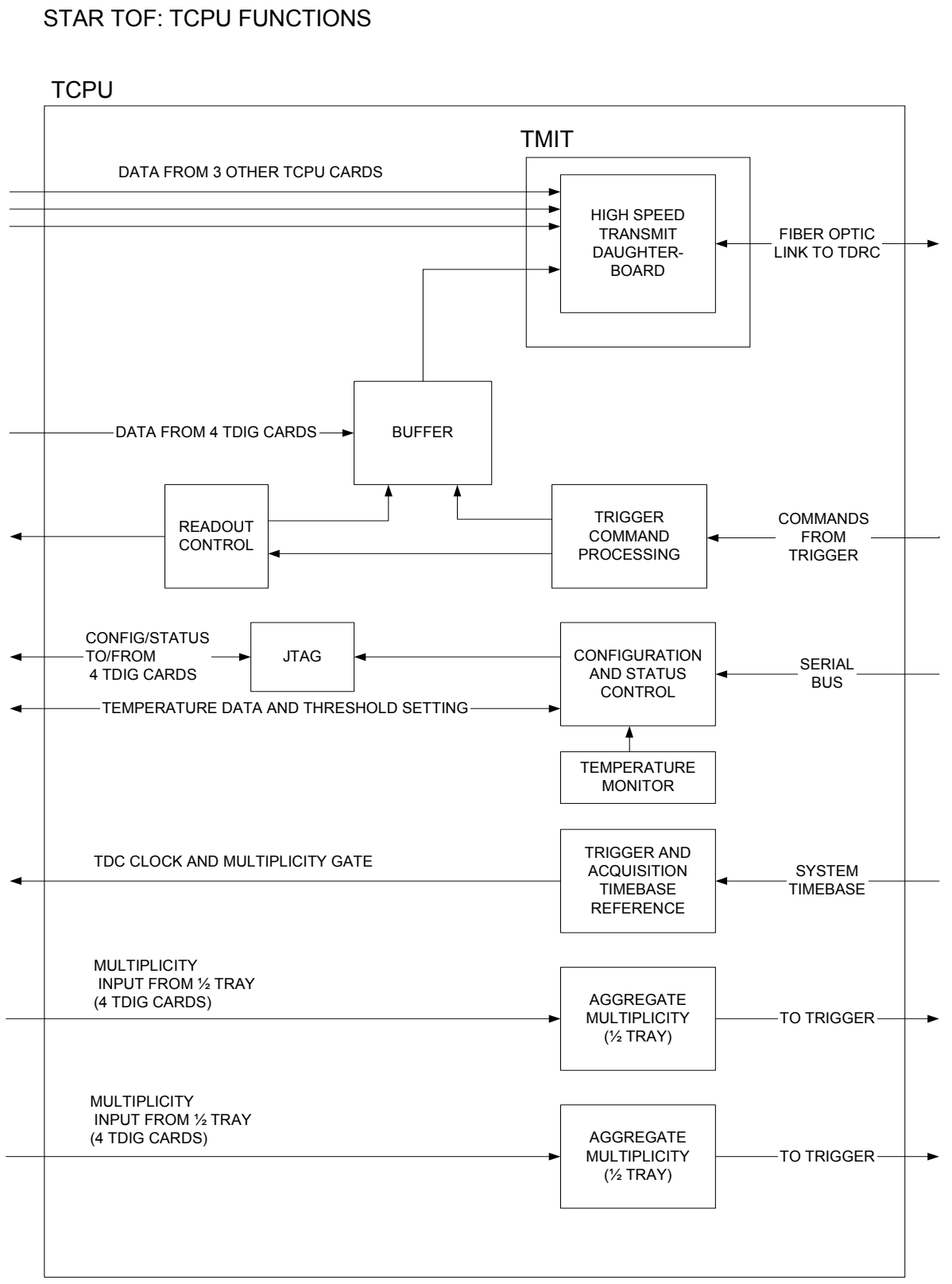

Figure 5: The tray level CPU (TCPU) and high speed data transmit (TMIT) circuit cards.

## 3.3. TCPU

The TCPU circuit card functions as a data concentrator and interface between the external experimental environment and the STAR TOF data acquisition electronics (TFEE and TDIG). The TCPU is implemented as a combination of embedded CPU and programmable logic. The TCPU functions at the detector tray level, and our design has one TCPU per tray. The TCPU performs the following functions (see Figure 5):

- Configure 8 pairs of TFEE/TDIG cards. Configuration parameters are loaded from on-board EEPROM storage or read from a host PC over the host serial bus. The TCPU sends configuration data to the TDC's over a JTAG daisy chain. The threshold DAC configuration data travels over a separate 2-wire communication path.
- Read and buffer data from all 8 TDIG cards of one tray (192 TDC channels).
- Aggregate and transmit 192 channels of multiplicity data to the STAR trigger as two 7-bit words representing the sum of all hit channels from 4 TDIG cards (96 detector channels). STAR trigger will use these multiplicity data in its Level-0 trigger decision.
- Maintain local data transmission between trays to combine tray level data (4 trays, unidirectional data flow).
- Use a TMIT daughter card (below) to communicate with STAR DAQ over a fiber optic link.
- Maintain a detector level local network (CAN bus) over up to 120 trays for configuration, calibration, integration, testing, monitoring, and maintenance.
- Receive and distribute a system level beam crossing clock from the RHIC accelerator. It conditions this signal by PLL frequency multiplication and filtering techniques to produce a TDC sampling clock with acceptably low jitter characteristics.
- Receive and process commands from the STAR trigger subsystem and initiate readout in response to those commands. It will also simply pass some commands such as trigger aborts and Level-2 trigger accepts on to DAQ via the TMIT card.

## 3.4. TMIT

The TMIT takes formatted data from the data/command buffer, serializes it and transmits it over a fiber optic link to the DAQ receiver "TDRC". The TMIT will be implemented as a PLD (or a communications chip specific to the communications protocol) and an optical transmitter. The TMIT is designed to be a flexible, easily replaced line unit. A change in the data communication environment will cause an upgrade only to the TMIT daughter board (30 cards in the system) and to field-upgradeable firmware on the TCPU board.

## 3.5. TDRC

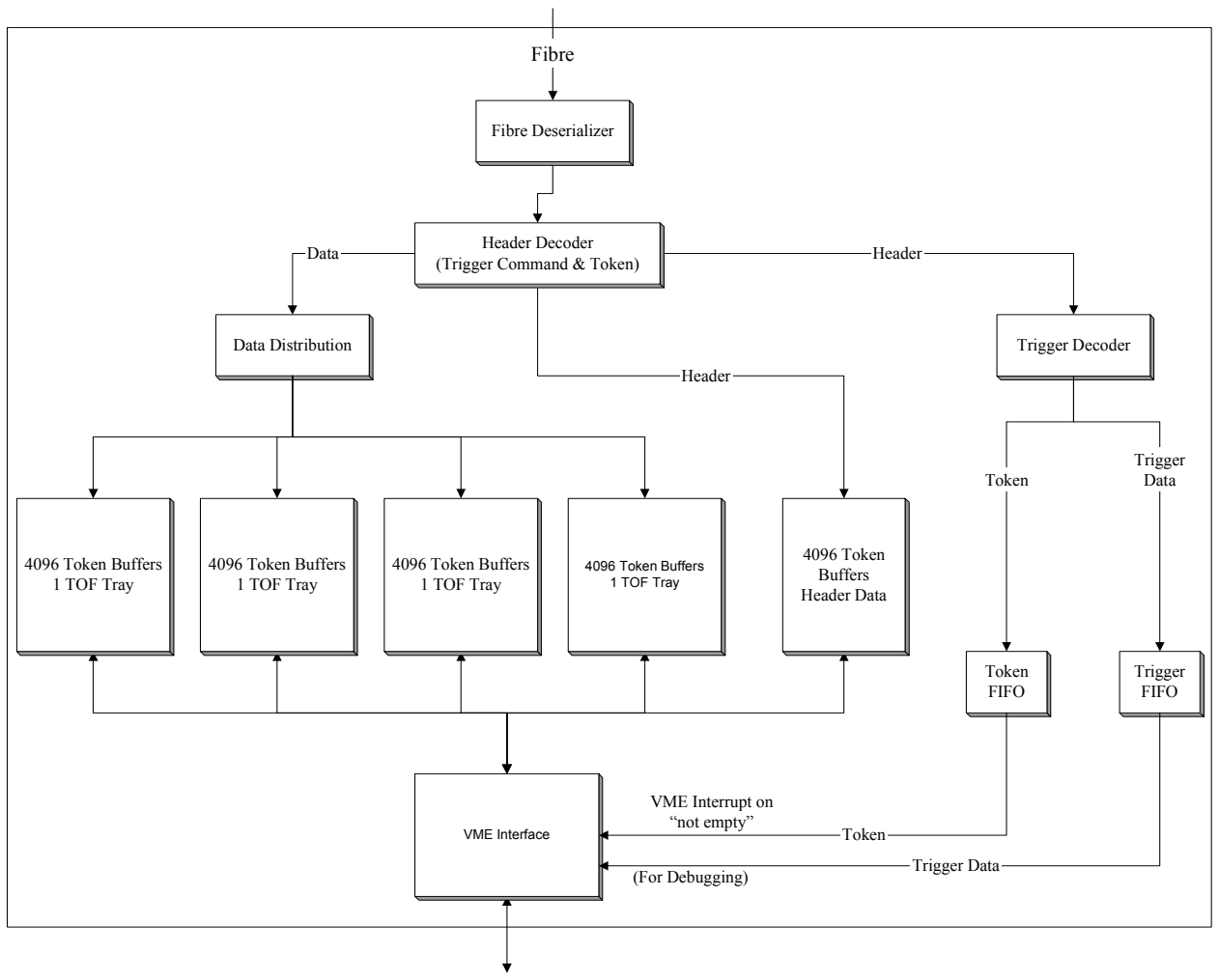

Figure 6: The principal components of the TDRC.

The TOF data acquisition receiver card ("TDRC") contains the optic link fiber receiver to receive data and trigger events from the on-detector fiber transmitter card TMIT. Each TDRC receives data from 4 TOF trays, resulting in a total of 30 TDRC cards for 120 TOF trays.

The principal components of the TDRC are shown in Figure 6. A fiber de-serializer receives the data from the optic link fiber and presents it as 20 bit words to a Header Decoder. The Header Decoder determines the type of event received. Fiber data consists of two possible event types: data events and trigger command events. On receipt of an event from the fiber, the TDRC decodes what kind of event it is and takes appropriate actions.

On receipt of a data event, the TDRC stores the associated data in one of a possible 4096 buffer locations locally, which correspond to the 4096 tokens possible from the STAR trigger system (the trigger tokens are implemented as 12 bit words). The addressing of these buffer locations is determined by the token number associated with the event. To simplify the buffer structure, the data is subdivided into data from each of the 4 TOF trays, which are stored in one of four buffer regions (possibly implemented as different mezzanine cards to the TDRC). The STAR trigger system assures that no two events in the system will have the same token associated with it. A token can only be reused by trigger after it has been returned to the trigger subsystem by DAQ, or an abort has been issued. This architecture guarantees that an event will never arrive at the TDRC with a token that is currently in use on the TDRC.

Each token memory location has associated with it an "invalidate" flag. Before data are written to a token memory location, the invalidate flag is checked to confirm that the previous content of this location is no longer needed. In case of a missing invalidate flag for a token, an error condition is indicated to VME.

On receipt of a trigger command event, the data are passed on to trigger decoder logic. Trigger command events contain the trigger command word, the DAQ command word, the trigger token, and possibly additional trigger information. Depending on the kind of trigger command word received, the buffer location associated with the token in the trigger command event will either be validated (in case of an L2-accept) or invalidated (in case of an abort). In case of a validated event, the associated token is pushed into a FIFO accessible from VME. A VME interrupt can be used to notify the processor in the VME crate of the arrival of a new event. The VME processor will then read the token from the FIFO, read the associated event fragments from the memory locations identified by the token, and invalidate the token memory locations (by setting the associated invalidate flag) so they can be reused. In case of an abort trigger event, the invalidate flag of the token memory location associated with the trigger token will be set, so that this location can be reused for future events. For debugging purposes, a second FIFO will be filled with a subset of the trigger information for each trigger received. This FIFO can also be accessed from VME to verify correct functioning of the trigger logic. For regular running conditions, this second FIFO can be disabled.

A VME interface allows read and write access to the different token memory regions, access to the token and trigger FIFO's, and handles interrupt generation and acknowledgement from VME.

## 4. PROTOTYPE TEST RESULTS

To verify that the goals of the above described design can be met, the TOF electronics group built prototype boards for both the TFEE and the TDIG cards.

An initial design of the TFEE board with 6 channels instead of the proposed 24 channels per board is currently employed on a prototype TOF tray in this year's RHIC run. This board has output provisions for both the digital signals and the analog signals, and is being read out with a CAMAC based system consisting of LeCroy TDC and ADC modules. Preliminary analysis of the data from the current RHIC run shows that the design meets the timing and noise requirements of the TOF proposal.

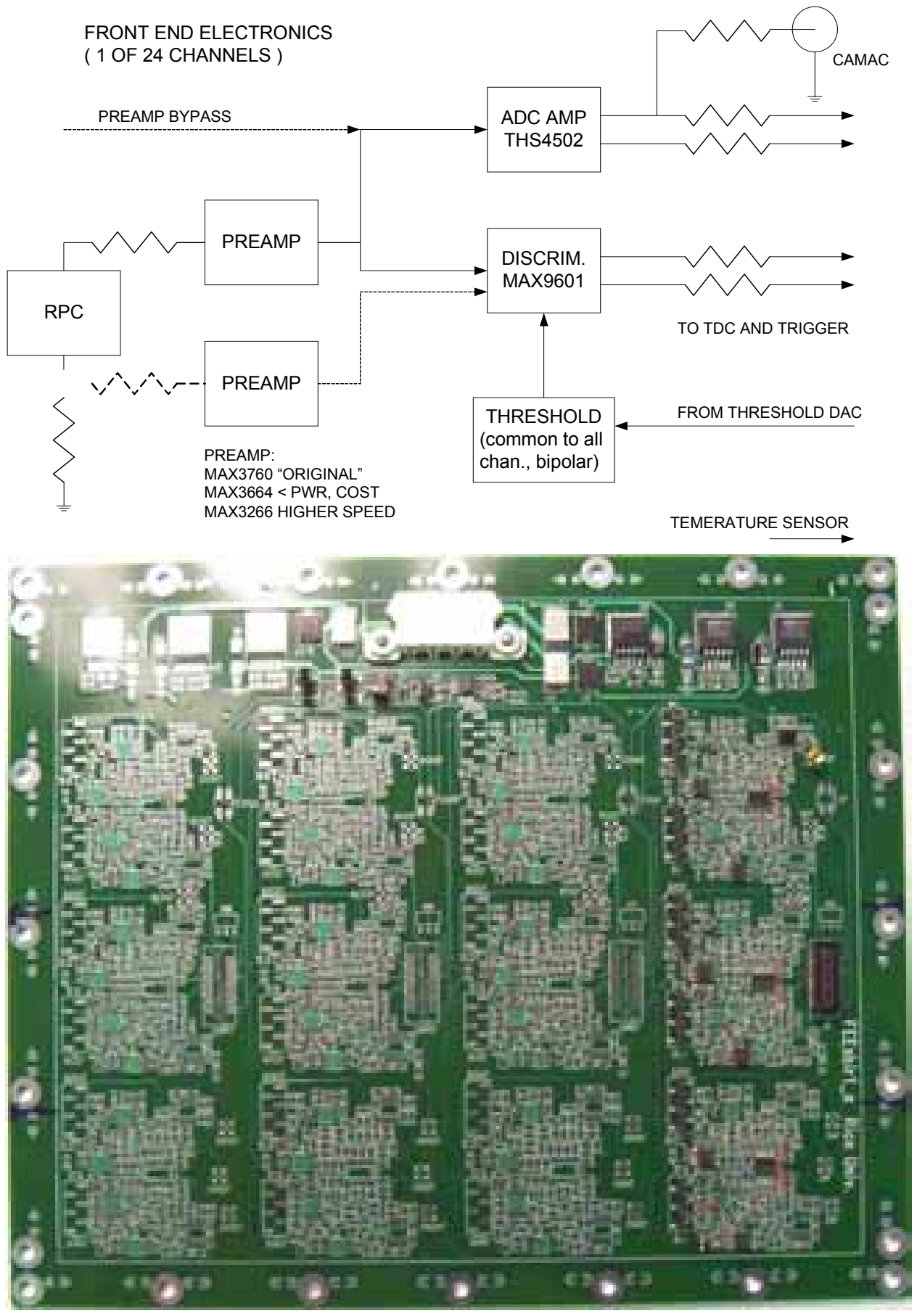

Figure 7: TFEE prototype block diagram and photo

Since its deployment in the prototype TOF tray, we have extended the design of this TFEE prototype to 24 channels, and adapted the form factor to match the final mechanical design of the MRPC system. We kept the analog output to be able to compare time walk corrections based on pulse height to corrections based on the time-over-threshold method proposed for the final system. On the 24 channel TFEE circuit board, we stuffed 6 of the 24 channels (one detector group) for testing. A block diagram of the schematics for one channel of this board and a photo of the prototype with 6 channels populated is shown in Figure 7. This circuit board required little troubleshooting. However, we did find that the low level input signals on this board and in our test setup were subject to RF interference, e.g. from people walking by with cell phones. We performed extensive stand-alone tests on the front end circuitry to verify low jitter, low crosstalk, and linear, repeatable performance. Integration of the analog electronics with the TDIG prototype board described below was essentially trouble free in a less noisy environment. Our tests connected the boards with a 6" length of ribbon coax cable that is plug compatible with the high speed connectors on each board (in the final design TDIG connects directly to TFEE with this connector). The analog circuitry performed very well in these tests.

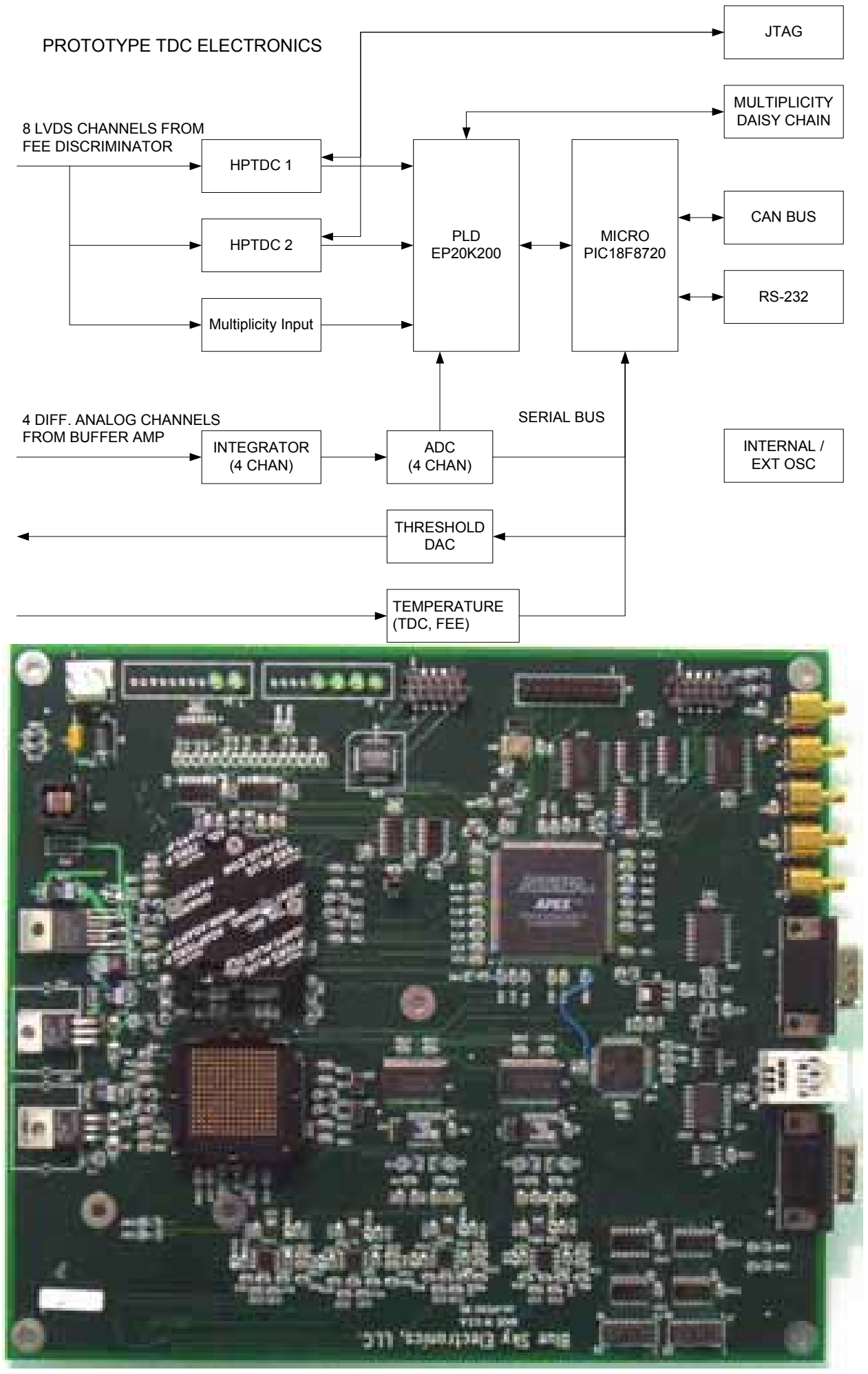

Figure 8: TDIG prototype block diagram and photo

The prototype of the TDIG board contains two CERN HPTDC chips to verify the timing resolution, chain readout, the time walk correction with time-over-threshold (by measuring both leading and trailing edges with the 2 TDC's), and ADC circuitry for pulse magnitude measurement to compare the different timewalk correction methods.

The TDIG prototype circuit board additionally contains a Microchip 18F8720 microprocessor. This processor is in place to manage board level configuration and communication. Hardware is in place on the board to enable intra-board communication over a CAN bus interface, for temperature monitoring, automatic TDC configuration and status, and automated discriminator threshold setting. The microprocessor and PLD are programmed in-circuit over removable serial cables, with no chip removal. A block diagram and a photo of this prototype are shown in Figure 8.

Testing of these prototypes is ongoing at this time. The main objectives of our tests with the TDIG prototype so far have been to understand the (rather complex) configuration of the CERN HPTDC chip and to study the timing resolution achievable with the HPTDC on TDIG alone and in combination with the TFEE prototype.

The HPTDC uses a tapped delay line architecture to perform its time-sampling function. The 40 MHz input clock is multiplied on the chip to 320 MHz with a phase locked loop (PLL). This 3.125 ns clock signal is fed through a 32-tap delay locked loop (DLL) which is sampled at the arrival of the hits, resulting in a measurement with 97.66 ps in the least significant bit (LSB). Additionally, the chip samples the input signal 4 times with a small time interval between the samples achieved with a 4-tap RC network. By determining in which sample the rising edge of the reference clock comes out of a DLL tap, one can deduct the arrival time of the hit with a resolution equal to the sample interval. This final subdivision results in bin widths of 24.41 ps. The dynamic range of the measurement is increased with the use of a clock-synchronous counter that is recorded together with each time sample. Each channel of the HPTDC has a small FIFO associated with it to buffer the data. Time measurements stored in the channel buffers are passed to a data processing unit and, after proper encoding, they are written into 256-deep latency buffers while waiting to be serviced by a trigger matching unit. The extraction of hits related to a trigger is based on trigger time tags within a programmable time window. Extracted measurements are stored in a common 256-deep readout FIFO until they are read out via either a parallel (8 or 32 bit) or a serial bus. The chip can also be operated in non-triggered mode, where every input hit is forwarded to the readout FIFO. Our initial tests with the prototype TDIG used the non-triggered mode of operation.

The HPTDC exhibits differential nonlinearity due to unequal bin widths in the RC tapped delay lines, unequal bin widths in the DLL, and noise coupling from the logic clock network (40 MHz, or 25ns) within the chip into the sampling clock network (320 MHz). The RC nonlinearity is periodic over 4 bins, the DLL nonlinearity is periodic over 32 bins, and the clock feed-through nonlinearity is periodic over 1024 bins (25ns). In all cases, these effects are deterministic and periodic for a given chip. As a result, these effects can be measured statistically and corrected for both on (in case of the DLL and RC nonlinearities) and off chip (in case of the clock feed-through). The designers of the HPTDC provided a set of on-chip calibration values for the DLL and RC nonlinearities that we found adequate in our tests.

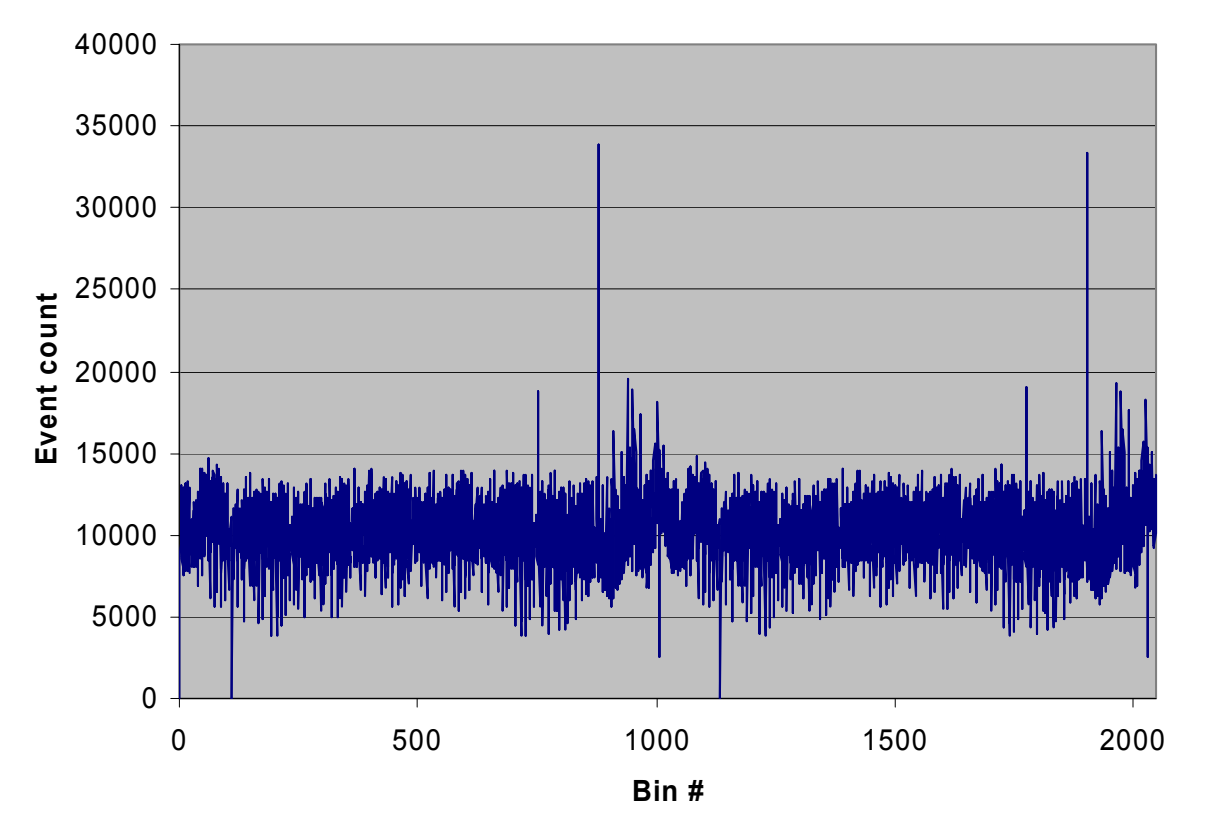

Figure 9: Code density results shown for 2048 bins

A standard technique for statistical bin width measurement is called a code density test [5]: a pulse generator running asynchronously with respect to the TDC produces pulses that arrive with a uniform random distribution with respect to the TDC clock. If all TDC bins were the same width, then the probability of a pulse arriving in any bin would be the same as that for any other bin, and a histogram of all bins would be flat (except for the statistical variation in the input distribution, which decreases with the number of hits per bin). Figure 9 shows the results of a code density test with 20 million pulse events acquired with CERN's recommended RC and DLL correction values. This plot clearly shows that the remaining differential non-linearity (DNL) from all sources is periodic over 1 HPTDC clock period (25 ns = 1024 bins). By integrating the DNL correction derived from this plot we arrived at the integral non-linearity (INL) shown in Figure 10. The values from this plot are used to construct a 10-bit INL correction table that was used by the analysis software to correct time difference measurements.

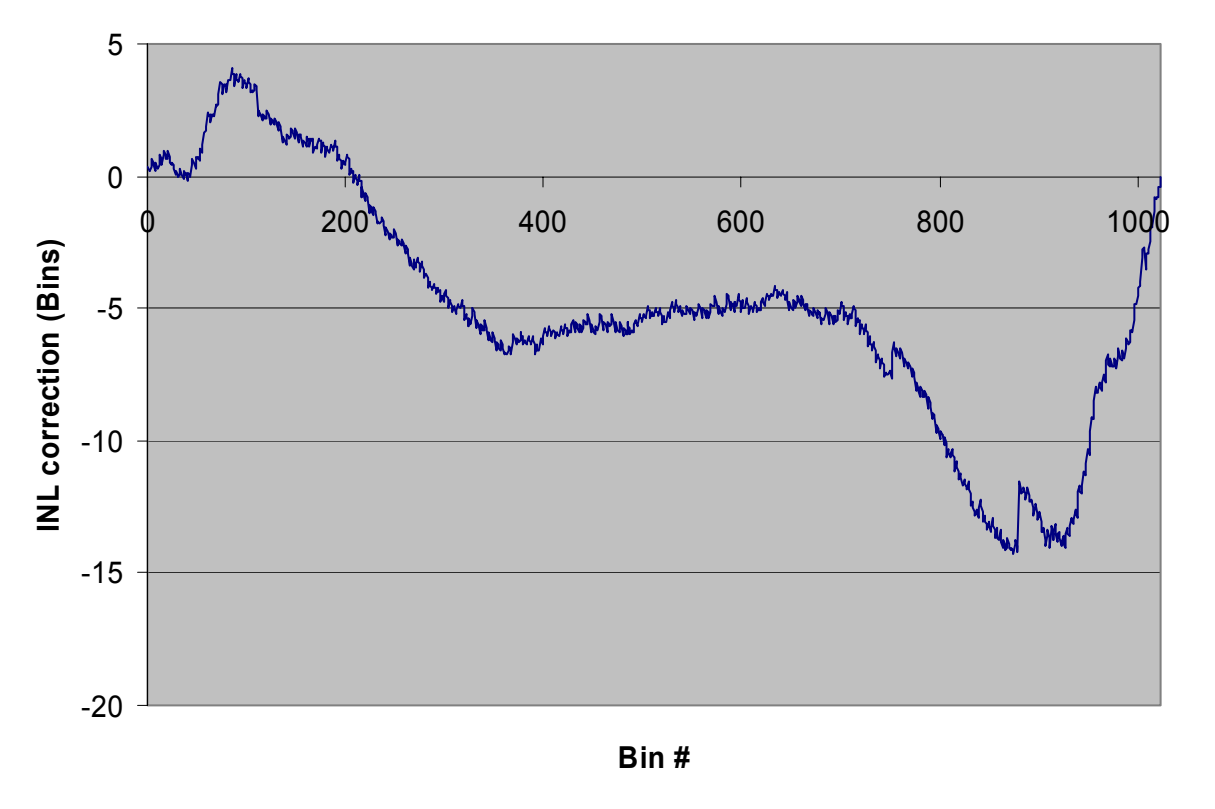

Figure 10: Integral non-linearity for the HPTDC

One of the possible reasons for the large values of the INL correction is that the HPTDC chips on the TDIG prototype are mounted in a BGA socket. Measurements from other groups that use the HPTDC showed that the socket adds an additional source of noise. We are able to mount the chip directly on the circuit board, and we will in the future investigate the effect of the socket. The reason for using the socket on TDIG was that we only had a limited amount of chips available for these tests.

To measure the time resolution of TDIG we used what is called a "cable delay test" by the HPTDC developers. In the cable delay setup we use a pulse generator that is free running with respect to the TDC clock. One output of the pulse generator is connected directly to one input of TDIG, while a second output is passed through a coaxial cable to generate a known and fixed delay before it is input into a second channel of TDIG. We then acquired pulse pairs in this setup and histogrammed the time difference of these pulses. Note that the individual time stamps measured in this setup will populate the whole dynamic range of the TDC, while the time difference should be constant. The result of this measurement is shown in the top histogram of Figure 10. The effect of the INL spreads the expected time difference peak and actually results in a double peak structure. However, after applying the INL correction described above, we arrive at the peak in the bottom histogram of Figure 10, which has an r.m.s. width of 32 ps. From this measurement one deduces a single channel r.m.s. of about 23 ps.

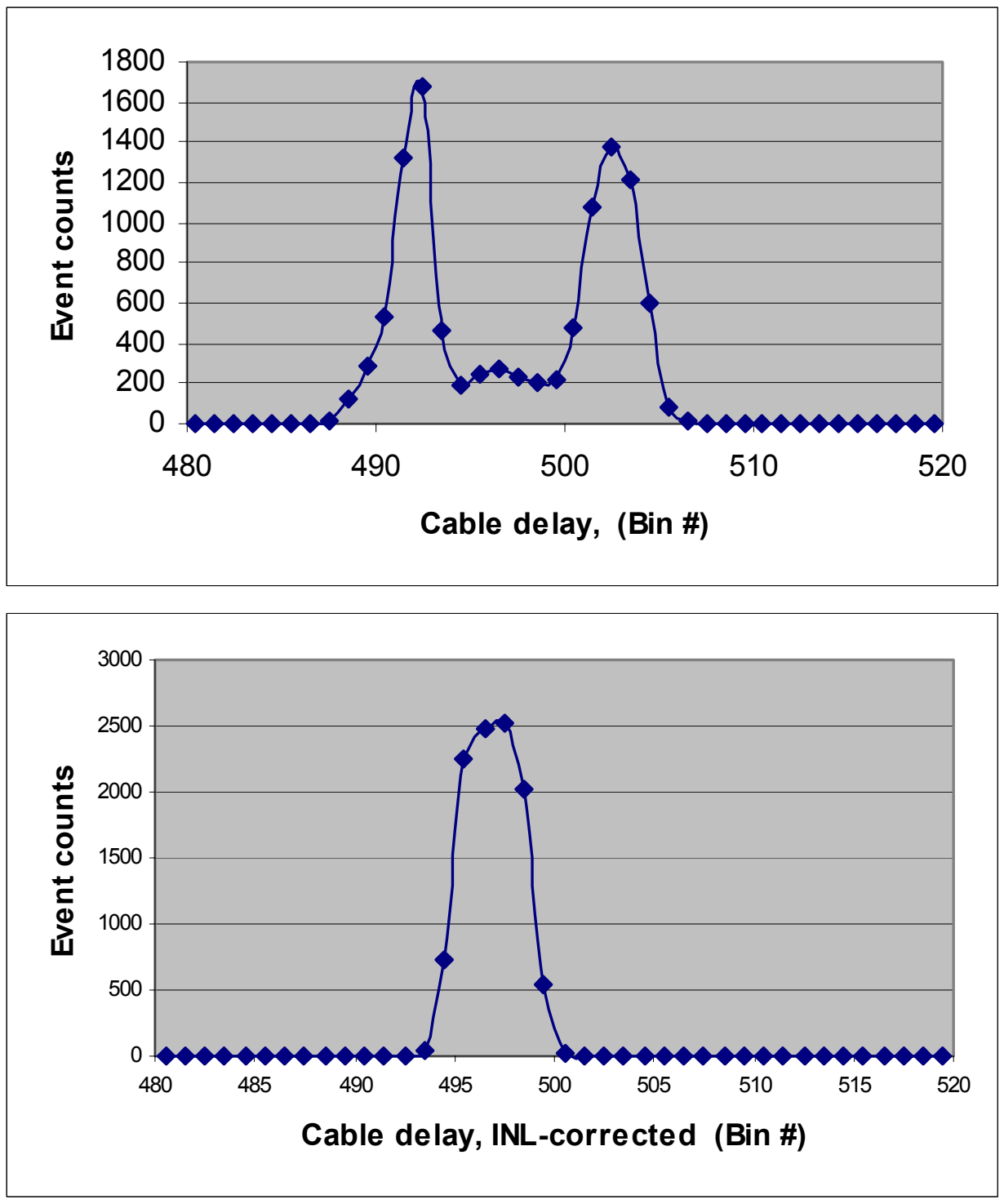

Figure 11: Cable delay data from TDIG alone, shown raw (top) and INL corrected (bottom histogram)

Figure 12 shows the r.m.s. peak widths, after INL correction, for a series of cable delays ranging over more than 2 times the INL period (25ns). The mean r.m.s. value of these measurements is 34 ps, implying a single channel rms of $34/\sqrt{2} = 24$ps.

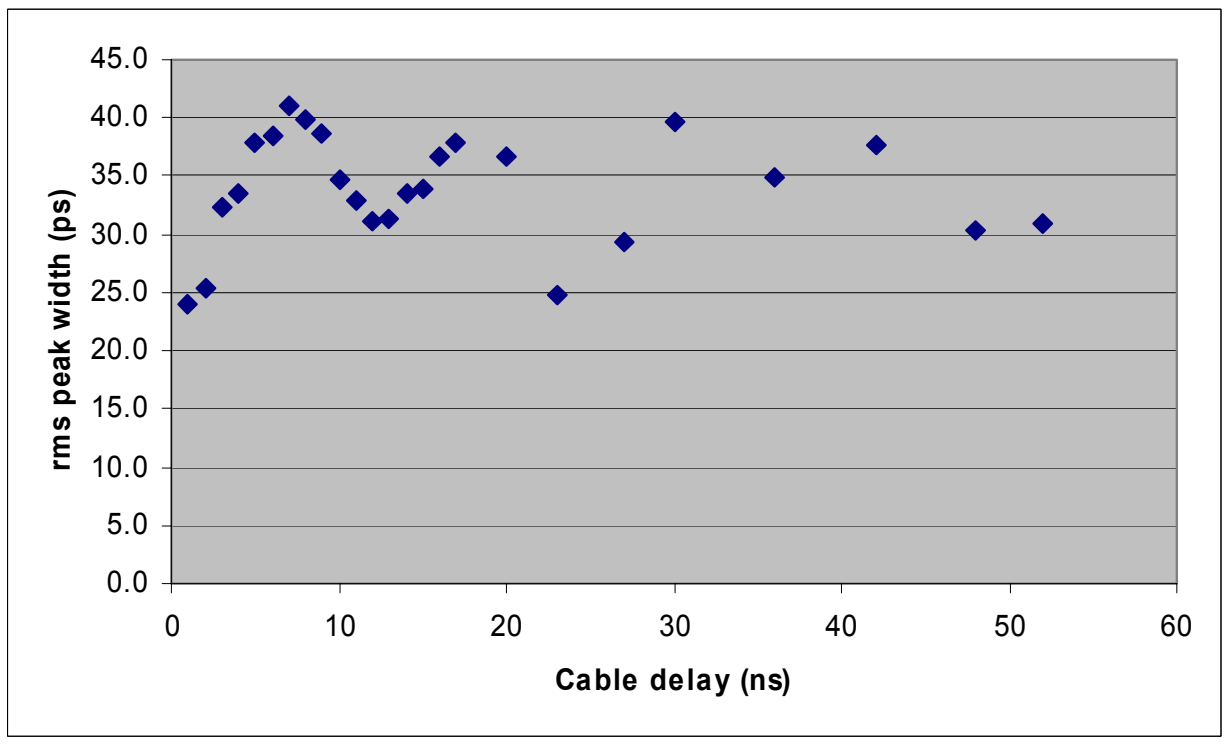

Figure 12: RMS peak width vs. cable delay, INL corrected

To test the effect of the whole electronics chain on timing resolution, we repeated this cable delay test with our analog front-end electronics. In this setup, we added one channel of our TFEE board (discriminator only) to the data path, after attenuation to the level of a photo multiplier tube output. This channel then emulates the "START" signal in the actual TOF detector setup. The second channel is delayed, attenuated, and converted to a current pulse by an analog test fixture. The discriminator thresholds are set at realistic levels determined with actual RPC detectors at the CERN test beam and in the current TOF prototype at the STAR experiment at RHIC. The attenuation factor is adjusted to 4 times the minimum value that triggers the discriminator. This pulse then goes through both amplifier and discriminator circuits in a second channel on the TFEE board, emulating a "STOP" signal in the actual MRPC detector setup. The resulting measurements are again recorded for different time intervals and INL corrected offline before they are histogrammed. An example histogram for this setup is shown in Figure 13.

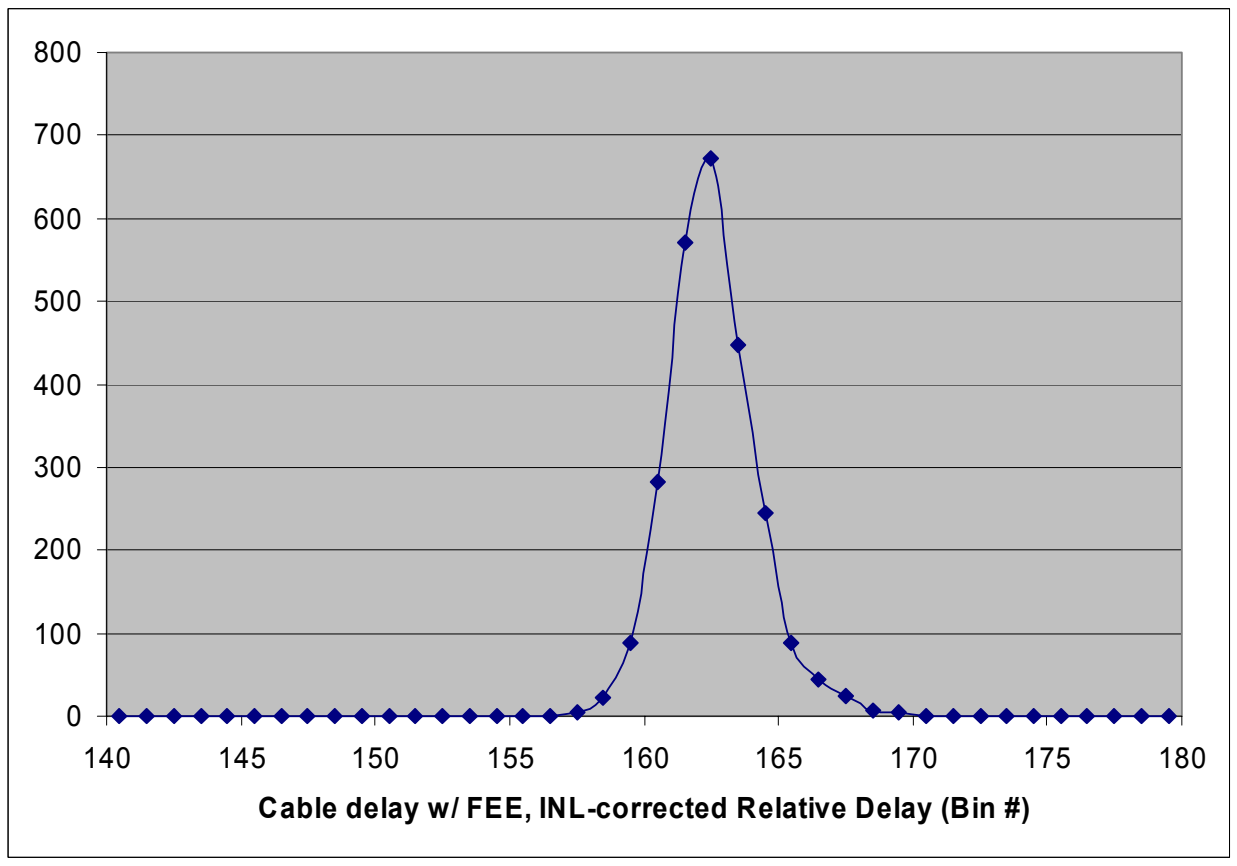

Figure 13: Cable delay data with front-end electronics, after INL correction

The mean of the rms peak width for several time delays was 42ps. The experimental setup in this test is very similar to the actual configuration in the TOF design, except that the analog signal will come from an MRPC detector and the time difference measurement will be performed in two different TDC's on different circuit boards.

## Acknowledgements

This work was supported in part by the U.S Department of Energy.